\documentclass[referee]{cjaa}           

\usepackage{graphicx}                   
\input{epsf.sty}                        
\input{psfig.sty}                       

\begin{document}

\title{Developing radio beam geometry and luminosity models of pulsars
} 


\author{P.L. Gonthier 
	\inst{1} \mailto{gonthier@hope.edu} 
	\and S.A. Story \inst{1} 
	\and B.M. Giacherio \inst{1} 
	\and R.A. Arevalo \inst{1}
\and A.K. Harding \inst{2} } 
\offprints{P.L. Gonthier}                   

\institute{Hope College, Department of Physics and Engineering, 27
Graves Place, Holland, MI 49424 USA\\ 
\email{gonthier@hope.edu} 
\and NASA Goddard Space Flight Center,
Laboratory for High Energy Astrophysics, Greenbelt, MD 20771 USA\\ 
}

\date{Received~~2005 month day; accepted~~2005~~month day}

\abstract{ Our recent studies of pulsar population statistics suggest
that improvements of radio and gamma-ray beam geometry and luminosity
models require further refinement. The goal of this
project is to constrain the viewing geometry for some radio pulsars,
especially three-peaked pulse profiles, in order to limit the
uncertainty of the magnetic inclination and impact angles. We perform
fits of the pulse profile and position angle sweep of radio pulsars for
the available frequencies. We assume a single core and conal beams
described by Gaussians. We incorporate three different size cones with
frequency dependence from the work of Mitra \& Deshpande (\cite{mitr99}). We
obtain separate spectral indices for the core and cone beams and explore
the trends of the ratio of core to cone peak fluxes. This ratio is
observed to have some dependence with period. However, we cannot
establish the suggested functional form of this ratio as indicated by
the work of Arzoumanian, Chernoff \& Cordes (\cite{arzo02}).
\keywords{radiation mechanisms:non-thermal - stars: magnetic fields - stars: neutron - pulsars: general} 
}

\authorrunning{P.L. Gonthier, S.A. Story, B.M. Giacherio, R.A. Arevalo
\& A.K. Harding }             

\titlerunning{Developing radio beam geometry and luminosity models of
pulsars }  


\maketitle 
%
%
\section{Introduction}           
\label{sect:intro}
A recent population synthesis study (Gonthier et al.~\cite{gont04}) that
incorporates the radio beam geometry and luminosity models of
Arzoumanian, Chernoff \& Cordes (\cite{arzo02}) (hereafter ACC) and the
geometry and luminosity of the polar cap slot gap model of Muslimov \&
Harding (\cite{musl03}) simulates the number of radio-loud and
radio-quiet (with a radio flux below the survey threshold) $\gamma$-ray
pulsars detected by the instruments EGRET, AGILE and GLAST.  The
simulated correlations between the radio and bright $\gamma$-ray beam profiles
suggest that $\gamma$-ray profiles with two peaks have a radio profile
that is core dominated with the core peak appearing in between the two
$\gamma$-ray peaks, which is contrary to the features of most of the
EGRET detected $\gamma$-ray pulsars that exhibit a single radio peak
leading in time the two $\gamma$-ray peaks (Thompson~\cite{thom01}).
Core-dominated radio emission is implied in the ACC model, especially for young $\gamma$-ray pulsars, due to
the assumption that the radio core-to-cone peak flux has a $1/P$ period
dependence.  These findings suggest that a better radio beam model is
needed to account for the observed correlations of radio and
$\gamma$-ray profiles.  Motivated by these results, we seek to develop
an alternate radio beam geometry and luminosity model by focusing on
well defined three-peak radio pulse profiles with adequate polarization
data.  We fit the polarization position angle within the
rotating vector model (RVM) of Radhakrishnan \& Cooke (\cite{radh69}) as
well as the pulse profiles. While we realize that other possible
interpretations of the structure of radio beams are possible, such as
the core being a conal structure (Kijak \& Gil \cite{kija02}) or the emitting surface being patchy
(Han \& Manchester \cite{han01}), we assume that pulsars are standard
candles, there is a single core beam and a single cone beam characterized
by an inner, middle or outer width following Mitra \& Deshpande
(\cite{mitr99}), and the rotating vector model is applicable.


\section{Model} 
\label{sect:Mod} 

The angular distribution of the core beam is assumed to be a Gaussian centered
along the magnetic axis with a characteristic width represented by
\begin{equation}\begin{array}{l} 
f_{{\rm core}} (\theta ) = \frac{1}{{\pi \rho _{\rm core}^2 }}e^{ - \theta ^2 /\rho _{\rm core}^2 }, \\ 
\rho _{{\rm{core}}}  = 1^ \circ  .5P^{ - 1/2} . \\
\end{array} \end{equation} 
where $P$ is the period in seconds and the factor in front of 
the exponent normalizes the Gaussian.  For a
given viewing geometry, defined by the inclination angle, $\alpha$, from
the magnetic axis and the viewing angle, $\zeta$, relative to the
rotational axis, then the polar angle, $\theta$, relative to the
magnetic axis is related to the phase angle, $\phi$, which is the
azimuthal angle relative to the rotational axis, through the expression
\begin{equation} 
\cos \theta  = \sin \alpha \sin \zeta \cos (\phi  - \phi _o ) + \cos \alpha \cos \zeta . 
\end{equation} 
The $\phi_o^{\rm core}$ is the
phase offset associated with the core beam of the pulsar profile.  The
characteristic width, $\rho_{\rm core}$, was adopted from ACC model.

The angular distribution of the emitting surface for the conal beam is
also parameterized with a Gaussian by the expressions
\begin{eqnarray} 
f_{\rm cone} (\theta ,\xi) &=& \frac{1}{{2\pi ^{3/2} w_e \bar\theta (\beta _{\rm ratio}  + 1) }}\left\{
{(\beta _{\rm ratio}  - 1) \left[ {1 - \sin \left( {\frac{\pi }{2}\sin
\xi } \right)} \right] + 2} \right\} e^{ - (\theta  - \bar \theta )^2/w_e^2 } {\rm{, where}} \nonumber \\ 
\rho _{{\rm{cone}}} & =& 4^ \circ  .8\left( {1 +
\frac{{66}}{\nu }} \right)P^{ - 1/2} , \nonumber \\ 
\bar \theta & =& {3\over 4}\delta _{\rm r} \rho _{\rm cone},{\rm{ and}} \nonumber \\ 
w_e  &=& \frac{{\delta _{\rm w} \rho _{{\rm{cone}}} }}{{4\sqrt {\ln 2} }},
\end{eqnarray}
where $\xi$ is the azimuthal angle in the magnetic frame and $\nu$ is the observing 
frequency in MHz.  The fitting parameter $\beta_{\rm ratio}$ is introduced to describe an
asymmetric conal distribution in order to have the ability to fit conal
peaks in the pulse profile that exhibit different
maxima.  The two fitting parameters, $\delta_{\rm r}$ and $\delta_{\rm w}$, allow
for various cones with $\delta_{\rm r}$=0.8, 1.0, and 1.3 for the inner,
middle, and outer cones of Mitra \& Deshpande (\cite{mitr99}).  While we
do not believe that the conal beam actually has this particular geometry
as opposed to say a patchy distribution, this function merely allows us
to describe the heights of the two conal peaks in the pulse profile with
a function that is easily integrated and, therefore, normalized.

We fit the position angle, $\psi$, of the polarization vector with the
RVM using the relationship
\begin{equation} 
\tan (\psi  - \psi^{PA}_o ) = \frac{{\sin \alpha
\sin (\phi  - \phi^{PA}_o )}} {{\sin \zeta \cos \alpha  - \cos \zeta \sin \alpha \cos (\phi  - \phi_o^{PA} )}}. 
\end{equation} 
The phase offset angle, $\phi^{PA}_o$ and the position angle offset,
$\psi_o$, can be obtained for each of the available frequencies, then
subtracted from the originals to obtain a global position angle sweep
for all frequencies.  The inclination angle $\alpha$ and the impact angle $\beta$ are related to each
other by the maximum rate of position angle  (PA) sweep $\psi_{\max}'$, via,
\begin{equation} 
\psi_{\max}'=\left( {\frac{{d\psi }}{{d\phi }}} \right)_{\max }  = \frac{{\sin \alpha }}{{\sin \beta }}.
\end{equation} 
Without high quality polarization data, it is not possible to uniquely determine the viewing
geometry, $\alpha$ and $\zeta$, from the position angle alone.  We
fit the global position angle to obtain the maximum rate of change 
of the position angle,$\psi_{\max}'$.  While the Stokes parameters I, Q, U and V have standard
deviations that obey Gaussian statistics, values derived form these
parameters do not. We follow the work of Everett \&
Weisberg (\cite{ever01}) in assigning errors to the position angle,
$\psi$.  Presented in Figure 1 is the global fit of the PA for pulsar, 
B2045-16.
\begin{figure}[ht]
   \begin{center}
   \mbox{\epsfxsize=0.9\textwidth\epsfysize=0.9\textwidth\epsfbox{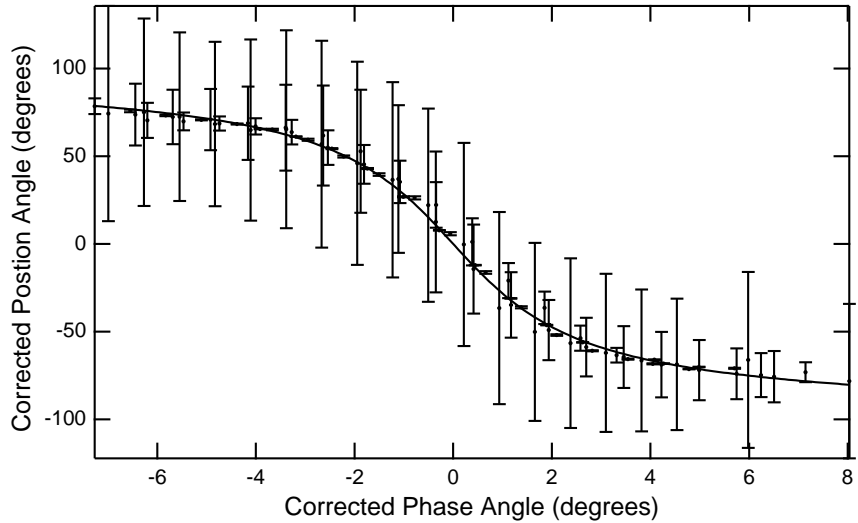}}
   \caption{The global position angle of the pulsar, B2045-16, that combines
   the data obtained of
   frequencies of 430, 606, 925, 1408 and 1624 MHz.  The smooth curve is the
   RVM fit.}
   \end{center}
\end{figure}

Assuming the maximum rate of change of the position angle $\psi_{\max}'$, we
fit the profiles constraining the inclination angle $\alpha$, and obtain the
impact angle, $\beta$.  Errors in the profile are determined by the standard deviation
of the off beam noise.  The fitting parameters associated with a
component of the profile are the overall amplitude, $A_{i}$, and
the phase offset, $\phi^{i}_o$, where $i$=core or cone.  The flux in the profile is
given by the expression
\begin{equation} 
s_i (\theta ,\nu ) = A_i(\nu) f_i(\theta,\nu ), 
\end{equation} 
where the $f_i$ term is defined above in Equations 1 and 3. 
Since the angular distributions are normalized over the emitting
surface, the coefficients, $A_i$,
represent the total angle integrated core and cone flux at the
particular frequency, $\nu$. However, the width, $\delta_{\rm w}$, and radius,
$\delta_{\rm r}$, of the conal annulus are allowed to vary to account for the
various cones suggested by the observations.  In addition, the
parameter, $\beta_{\rm ratio}$ allows for the two peaks of the conal
beam in the pulse profile to have different contributions.  Assuming a
power law with a low frequency cutoff of 50 MHz, the frequency-differential
flux has the form
\begin{equation}
s_i(\nu ) =  - \frac{{\alpha_i  + 1}}{\nu }\left( {\frac{\nu }{{{\rm{50 MHz}}}}} \right)^{\alpha_i  + 1} \frac{L_i}{{d^2 }},
\end{equation}
where $i=$ core or cone, $\alpha_i$ is the spectral index, $d$, is the
assumed distance to the pulsars and $L_i$ is the angle and frequency
integrated luminosity.  Having the amplitudes as a function of the
available frequencies, the spectrum can be fit with this function to
obtain the luminosity and spectral index of the core and conal components, separately.

Since the maximum rate of change of the PA, $\psi_{\max}'$, is the slope of the sweep at the
inflection point of the curve, the minimum $\chi^2$ of the RVM fit to the PA is weakly
dependent on the inclination angle, $\alpha$, as seen in Figure 2(a), where the 
$\chi^2$ of the fit is plotted as a function of the maximum rate of change in the PA, $\psi_{\max}'$,
and $\alpha$.  On the other hand, the $\chi^2$ obtained from fitting the profile is
weakly dependent on the maximum rate of change of the PA, $\psi_{\max}'$, as seen in Figure 2(b).  Therefore,
fitting both the PA and the profile gives a better constraint of the viewing geometry.
\begin{figure}[ht]
   \begin{center}
   \mbox{\epsfxsize=1.\textwidth\epsfysize=1.\textwidth\epsfbox{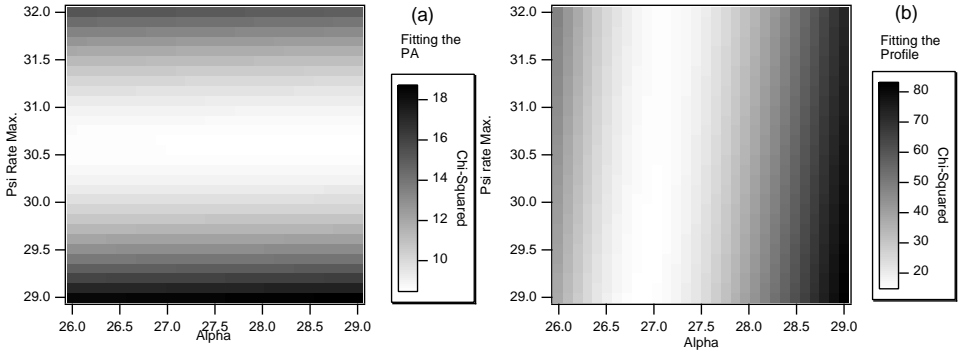}}
   \caption{The reduced $\chi^2$ obtained from fitting the PA (a) and the profile (b)
   of the pulsar B2045-16
   as a function of the maximum rate of change of the PA, $\psi_{\max}'$, and the inclination angle, $\alpha$,
   at a frequency of 430 MHz.}
   \end{center}
\end{figure}

The largest uncertainty of the procedure is introduced by the selection of the
radius of the annulus parameter, $\delta_{\rm r}$, which defines the inner, middle or outer
cone.  Values of $\delta_{\rm r}=0.8, 1.0$ or $1.3$ result in different inclination
angles.  For example in the case of the pulsar, B2045-16, once the maximum rate of change of the PA
is established and allowing all parameters to vary while fitting the profile, the minimum $\chi^2=294$
is obtained for $\delta_{\rm r}=1.00$, $\delta_{\rm w}=0.65$ and $\alpha=27^\circ.9$, suggesting the preference for the middle
cone.  However, as can be seen in Table 1, there is a large variation in the values of $\alpha$ with the 
conal radius parameter, $\delta_{\rm r}$. 
\begin{table}[ht]
  \caption[]{ Averages and standard deviations of the parameters obtained from fitting the profiles of the pulsar, B2045-16,
  for the inner, middle and outer cones at the frequencies of 408, 606, 925, 1408 and 1642 MHz.}
  \label{Tab:fitsB2045}
  \begin{center}\begin{tabular}{lcccccc}
  \hline\noalign{\smallskip}
Cone &  $\delta_{\rm r}$      & $\delta_{\rm w}$ & $\beta_{\rm ratio}$ & $\alpha$ & $\beta$ & $\chi^2$             \\
  \hline\noalign{\smallskip}
  inner & 0.8  & $0.50\pm 0.03$ & $0.76\pm 0.13$ & $21.7\pm 0.6^\circ$ & $-0.67\pm 0.02^\circ$ & $356\pm 410$ \\ 
  middle & 1.0 & $0.65\pm 0.04$ & $0.69\pm 0.10$ & $27.8\pm 0.7^\circ$ & $-0.84\pm 0.03^\circ$ & $265\pm 293$ \\
  outer  & 1.3 & $0.93\pm 0.08$ & $0.65\pm 0.09$ & $37.6\pm 0.9^\circ$ & $-1.10\pm 0.04^\circ$ & $544\pm 706$ \\
  \noalign{\smallskip}\hline
  \end{tabular}\end{center}
\end{table}
Based on the average minimum $\chi^2$ of 265, the middle cone is selected, and we show in Figure 3
the resulting profiles and fits of the pulsar, B2045-16, for the indicated frequencies.
\begin{figure}[ht]
   \begin{center}
   \mbox{\epsfxsize=0.9\textwidth\epsfysize=0.9\textwidth\epsfbox{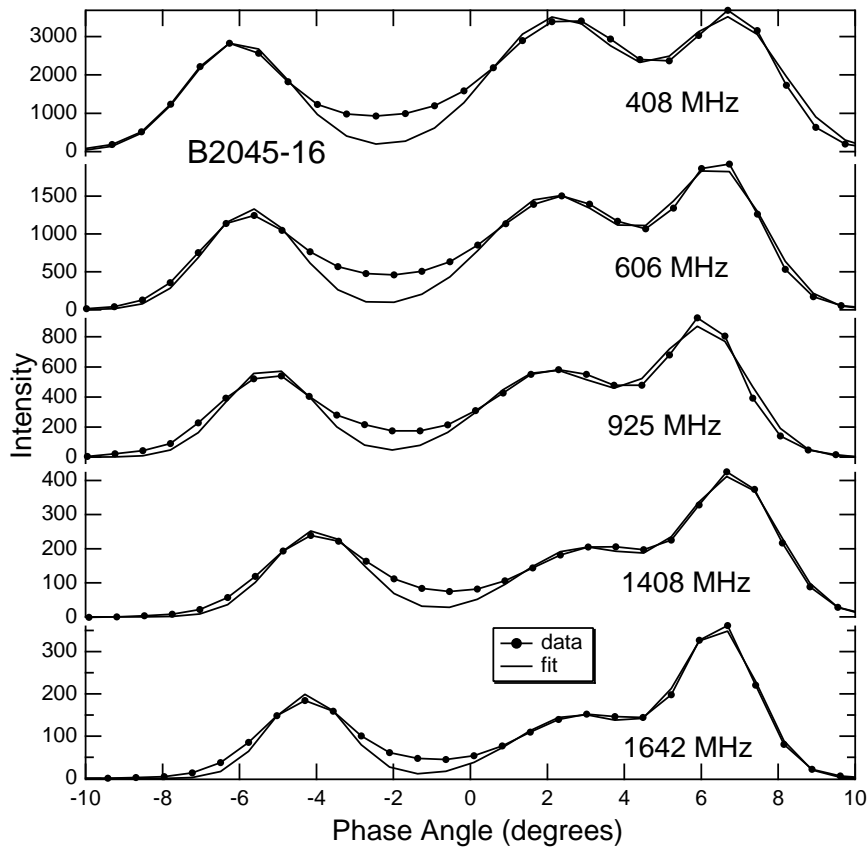}}
   \caption{The pulse profiles (curves with dots) and the fits (smooth curves) of the
   pulsar B2045-16 at the indicated frequencies.}
   \end{center}
\end{figure}

In Figure 4, we plot the core and cone coefficients, $A_{\rm core}$ and
$A_{\rm cone}$, representing the angle integrated flux of Equation 6 from the
fits of the profiles of pulsar, B2045-16, as a function of the available
frequencies indicated.  The smooth curves are fits of the power law
given in Equation 7, resulting in the indicated total angle and frequency
integrated fluxes and spectral indices for the core and cone beams.
\begin{figure}[ht]
   \begin{center}
   \mbox{\epsfxsize=0.8\textwidth\epsfysize=0.8\textwidth\epsfbox{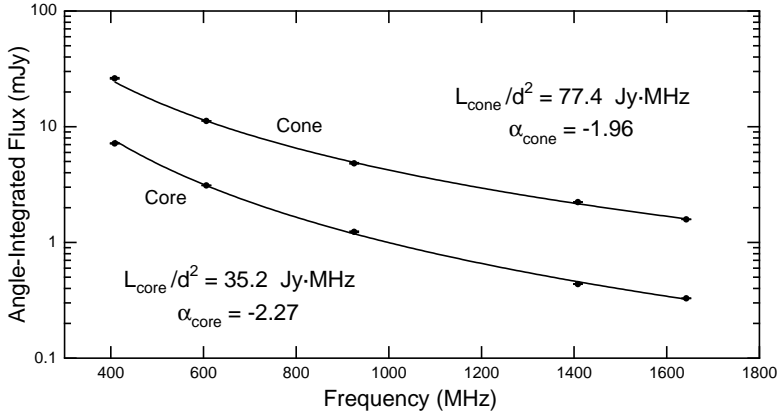}}
   \caption{The core and cone angle integrated fluxes obtained from the fits to the profiles
   of the pulsar B2045-16 at the frequencies of 408, 606, 925, 1408, and 1642 MHz.  Assuming
   the power law in Equation 7, the fluxes are fit to obtain the indicated angle and frequency
   integrated fluxes and spectral indices.}
   \end{center}
\end{figure}
Using the distances (with $\approx$20\% uncertainty) of the pulsars from the new 
distance model of Cordes \& Lazio (\cite{cord02}), we can estimate the core and cone luminosities.
\section{Results} 
\label{sect:results} 

We selected a group of seventeen pulsars ranging in period from 0.06 s
(B1913+16) to 3.7 s (B0525+21) whose pulse profiles exhibit a fairly well
defined structure consisting of three peaks and have adequate
polarization data.  We obtained the data from the European Pulsar
Database (EPN) with all profiles and polarization data taken in the survey by
Gould \& Lyne (\cite{goul98}). From the fitting of the position angle
sweep and pulse profiles, we obtain the inclination angles, $\alpha$,
the impact angle, $\beta$, the spectral indices, fluxes and luminosities
of the core and cone beams separately. In Table 2, we compare the
extracted inclination $\alpha$, impact angles $\beta$ and the maximum rate of change of
the PA, $\psi_{\max}'$, with the works of Lyne \& Manchester
(\cite{lyne88}) and Rankin (\cite{rank93}). With this information, we
are able to estimate the ratio of the core-to-cone peak fluxes, which is
independent of the pulsar distance and associated uncertainties.   In
Figure 5, we present this ratio as a function of the period for three
different indicated frequencies.  The predictions of the ACC model that
assume the ratio of core-to-cone peak flux has a $1/P$ dependence are
indicated by dotted lines in each of the panels.
\begin{table}[ht]
  \caption[]{ Inclination angles $\alpha$, impact angles $\beta$, and maximum rate of change of PA $\psi_{max}'$ obtained
  from fits are compared to the studies of Lyne \& Manchester (\cite{lyne88}) and Rankin (\cite{rank93}).
  In the fits $\alpha\leq 90^\circ$ and $\beta$ and $\psi_{max}'$ have the same signs.} 
  \label{Tab:Angles}
  \begin{center}\begin{tabular}{lccccccccc}
  \hline\noalign{\smallskip}
Pulsar & \multicolumn{3}{c}{This Work}  & \multicolumn{3}{c}{Lyne \& Manchester} & \multicolumn{3}{c}{Rankin} \\
  \hline\noalign{\smallskip}
Name & $\alpha$ & $\beta$ & $\psi_{\max}'$ & $\alpha$ & $\beta$ & $\psi_{\max}'$ & $\alpha$ & $\beta$ & $\psi_{\max}'$ \\
  \hline\hline\noalign{\smallskip}
B1913+16 & 44 & -0.8 & -52 &     &      &     & 46 & -0.8 & -51 \\
B1804-08 & 51 & 3.5  & 13  & 47  & 2    & 21  & 63 & 5.1  & 10  \\
B1702-19 & 45 & 5.7  & 7   & 90  & 4.1  & 14  & 85 & 4.1  & 14  \\
B2048-72 & 35 & -2.3 & -14 & 29  & -1.4 & -20 &    &      &     \\
B1839+09 & 90 & 2.6  & 22  & 90  & 2.9  & 20  & 83 & 1.4  & 42  \\
B0727-18 & 25 & -4   & -7  & 28  & -6.8 & -4  &    &      &     \\
B2003-08 & 19 & -3.7 & -5  & 13  & -3.1 & -4  & 13 & -3.3 & -4  \\
B1845-01 & 42 & -3.2 & -12 &     &      &     &    &      &     \\
B1508+55 & 35 & 2.9  & 11  & 80  & 2    & 28  & 45 & 2.7  & 15  \\
B1821+05 & 27 & 2.9  & 10  & 28  & 1.5  & 18  & 32 & 1.7  & 18  \\
B2111+46 & 10 & 1.1  & 9.1 & 8.6 & 1.3  & 6.7 & 9  & 1.4  & 6.7 \\
B1039-19 & 24 & 1.3  & 18  & 34  & 1.8  & 18  & 31 & 1.7  & 18  \\
B2045-16 & 27 & 0.8  & 31  & 37  & 1.1  & 30  & 36 & 1.1  & 30  \\
B0525+21 & 14 & -0.4 & -35 & 23  & -0.7 & -31 & 21 & -0.6 & -36 \\
  \noalign{\smallskip}\hline
  \end{tabular}\end{center}
\end{table}
\begin{figure}[ht]
   \begin{center}
   \mbox{\epsfxsize=1.\textwidth\epsfysize=1.\textwidth\epsfbox{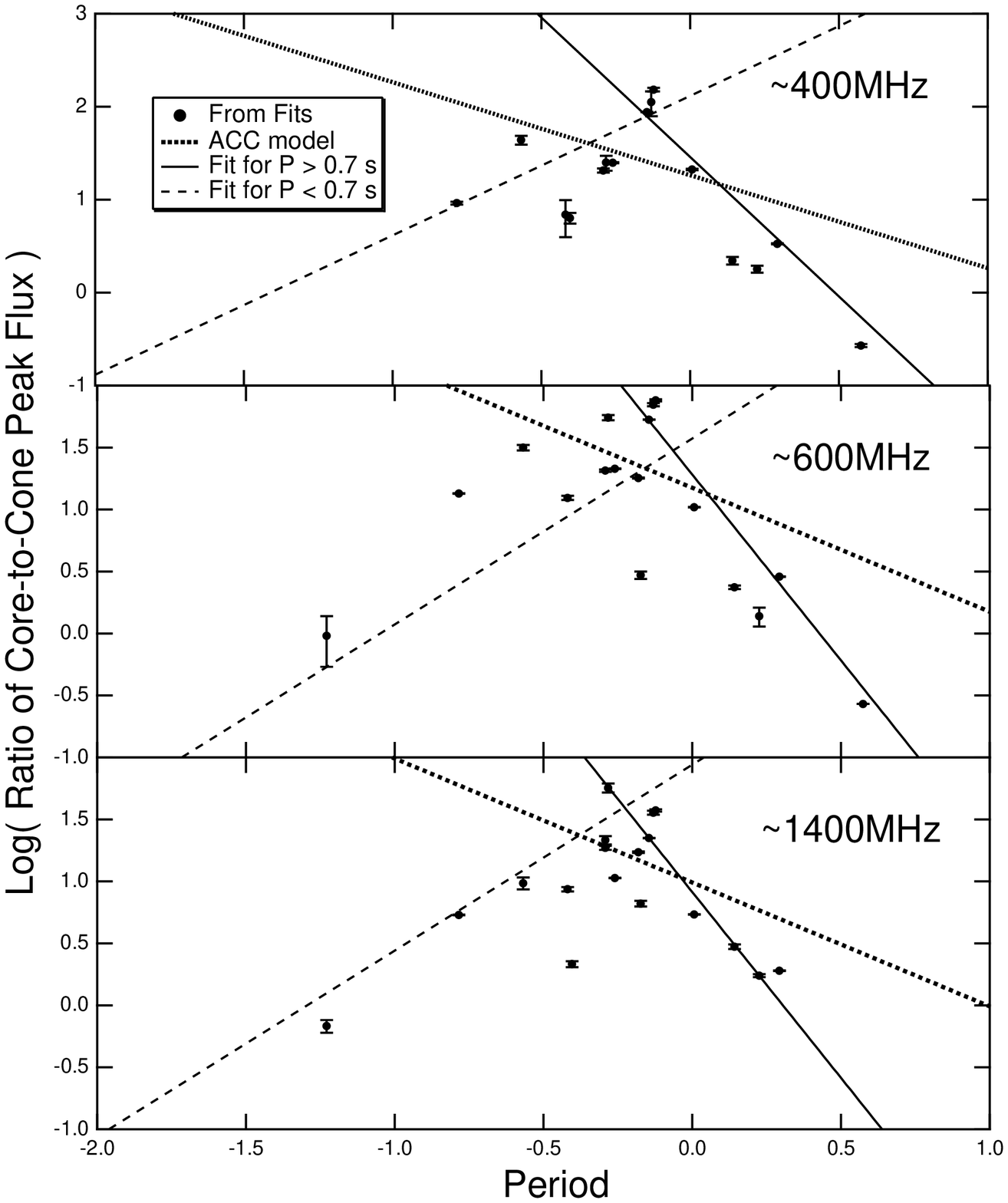}}
   \caption{The logarithm of the ratio of the core-to-cone peak flux is plotted
   as a function of the logarithm of the pulsar period for the indicated frequencies.  
   The predictions of the ACC model are indicated by dotted lines.  The dashed and
   solid lines represents the
   parameterization of the ratios below and above 0.7 seconds, respectively (see text).}
   \end{center}
\end{figure}
The period dependence of the ratios indicates a break around a
period of 0.7 seconds.  We parameterize the period and frequency
dependence of the ratios below and above 0.7 s.  We find that the ratio
of the core-to-cone increases as $P^{1.5}\nu^{-1}$ up to 0.7 seconds and
then decreases as $P^{-3}\nu^{-1}$. The results are quite different
than those of the ACC model, particularly for the behavior of the ratio
core-to-cone peak flux for short period pulsars. Our results suggest
that they are not as core-dominated as implied by an extrapolation of
the ACC model.

With the extracted fit parameters, we are able to estimate the
luminosity of the core and cone beams separately, as shown in Figure 6.  The
bottom part of the figures displays the luminosity of the cone beam
(solid dots) with a parameterized function suggested by the fits
(crossed circles).  The functional dependence of period and period
derivative is indicated in the figure for the parameterization. Likewise
the luminosity of the core is displayed in the middle portion of the
figure.  The total luminosity is shown at the top of the figure along
with the predictions of the ACC model.
\begin{figure}[ht]
   \begin{center}
   \mbox{\epsfxsize=1.\textwidth\epsfysize=1.\textwidth\epsfbox{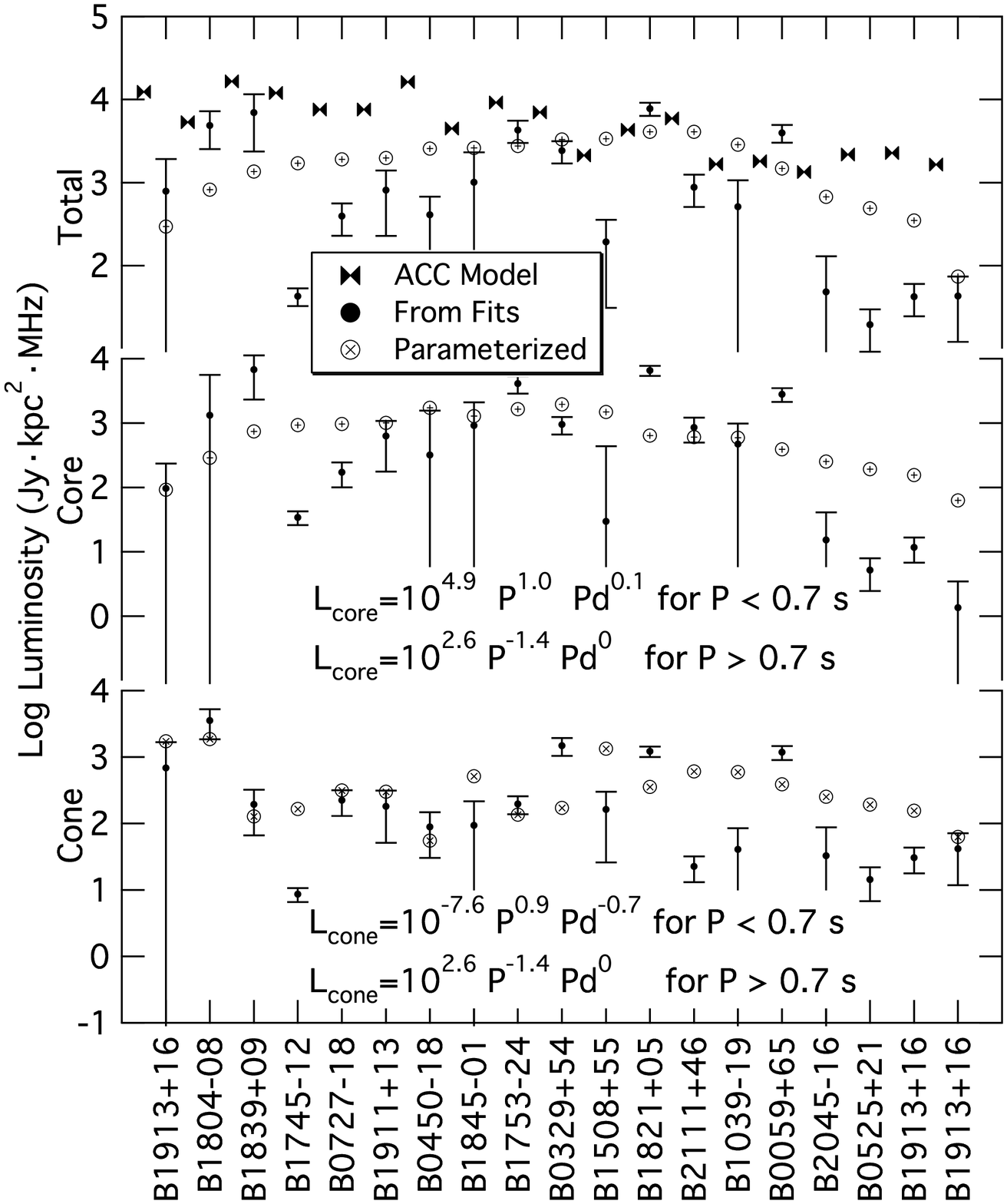}}
   \caption{The logarithm of the total (top), core (middle) and cone (bottom) luminosities
   obtained from the fits (solid dots), parameterization (circles with crosses) and predictions of ACC model (top).}
   \end{center}
\end{figure}

\section{Conclusions} 
\label{sect:conclusion} 

We focus on a select group of pulsars that exhibit three peaks in their
pulse profiles with adequate polarization data.  We find that there are
a very limited number of these pulsars on the EPN database with good polarization data. 
All of polarization data in this study comes from the survey of Gould \& Lyne
(\cite{goul98}).  We fit both the position angle as well as the
pulse profile intensities with a simple model that assumes a single core
beam and a single cone beam.  Assuming a standard period dependence of
the characteristic widths (Equations 1 and 3) and the presence of three conal beams as
described by Mitra \& Deshpande (\cite{mitr99}), we find that the ratio
of the core-to-cone peak flux for short period pulsars does not follow
the $1/P$ trend suggested by the study of Arzoumanian, Chernoff \&
Cordes (\cite{arzo02}). There appears to be a break in the period
dependence of the ratio near a period of 0.7 seconds, with the ratio
decreasing for short period pulsars.  Such a result alters significantly
the interpretation of radio profiles of short period pulsars and
requires further study.

The interpretation of radio profiles in young high-energy pulsars is not
well established. While Arzoumanian, Chernoff \& Cordes (\cite{arzo02})
proposed both a beam geometry and a luminosity model of the core and
cone radio beams,  a considerable controversy exists in the
interpretation of single peaked radio pulse profiles, as to whether
the origin is from core emission or partial cones (Manchester
\cite{mans01}).  The study presented in this paper and others like it
may help in understanding the geometry of radio emission from pulsars,
particularly from young pulsars.  Our study indicates that the ACC model
requires some revision in regards to the ratio of the core-to-cone
emission for short period pulsars.

\begin{acknowledgements} We express our gratitude for the generous
support of the Michigan Space Grant Consortium, of Research Corporation
(CC5813), of the National Science Foundation (REU and AST-0307365) and
the NASA Astrophysics Theory Program. 
\end{acknowledgements}

\label{lastpage}

\end{document}